\documentclass{PoS}

\def\gsim{\mathrel{\rlap{\lower4pt\hbox{\hskip1pt$\sim$}}
    \raise1pt\hbox{$>$}}}         %greater than or approx. symbol
\def\lsim{\mathrel{\rlap{\lower4pt\hbox{\hskip1pt$\sim$}}
    \raise1pt\hbox{$<$}}}         %less than or approx. symbol

\title{Progress in the NNPDF global analysis and the impact of the legacy HERA combination}

\ShortTitle{Progress in the NNPDF global analysis}

\author{\speaker{Juan Rojo}\thanks{On behalf of the NNPDF Collaboration.}\\
        Rudolf Peierls Centre for Theoretical Physics, 1 Keble Road,\\
University of Oxford, OX1 3NP Oxford, United Kingdom\\
        E-mail: \email{juan.rojo@physics.ox.ac.uk}}

\abstract{
The H1 and ZEUS collaborations have recently
  presented their final results for the combination of
  inclusive cross-section measurements taken during
  Run I and Run II at the HERA collider.
  In this contribution, following an overview of recent progress
  in the NNPDF framework,
  we quantify the impact of
  the legacy HERA dataset on the
  NNPDF3.0 analysis, finding that it has
  a very moderate effect in the global fit.
  On the other hand, we also find that a HERA-only fit using the legacy dataset leads
  to a rather more accurate determination of PDFs as compared to a fit including only
  the HERA-I data.
  We also explore the sensitivity of the fit with respect
  to kinematical cuts in the small-$x$ and small-$Q^2$ region,
  finding hints of a possible tension between data and theory.
}

\FullConference{The European Physical Society Conference on High Energy Physics\\
		22--29 July 2015\\
		Vienna, Austria}

\begin{document}

\paragraph{PDFs for the LHC Run II.}

\begin{figure}[t]
\centering
\includegraphics[width=0.49\textwidth]{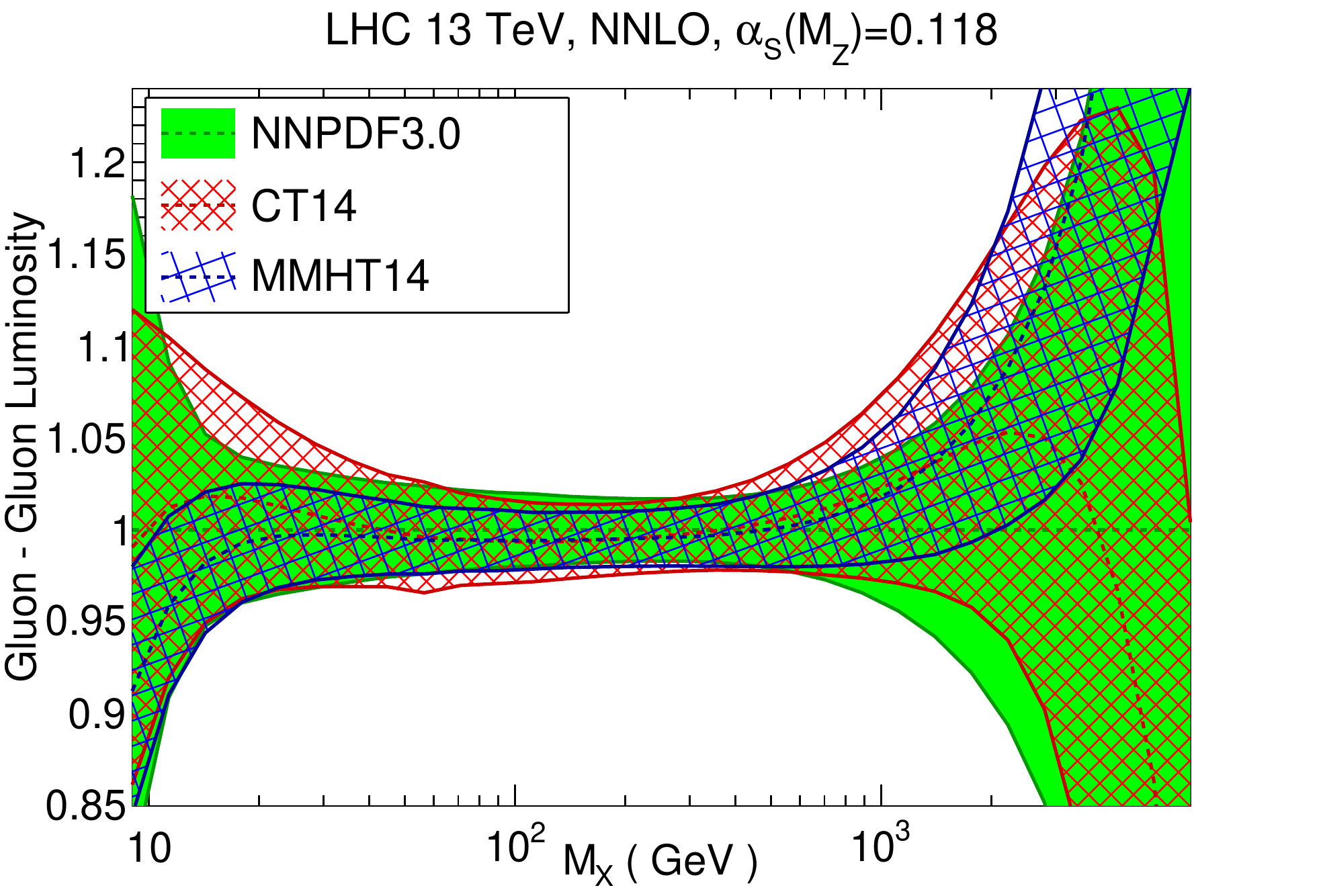}
\includegraphics[width=0.49\textwidth]{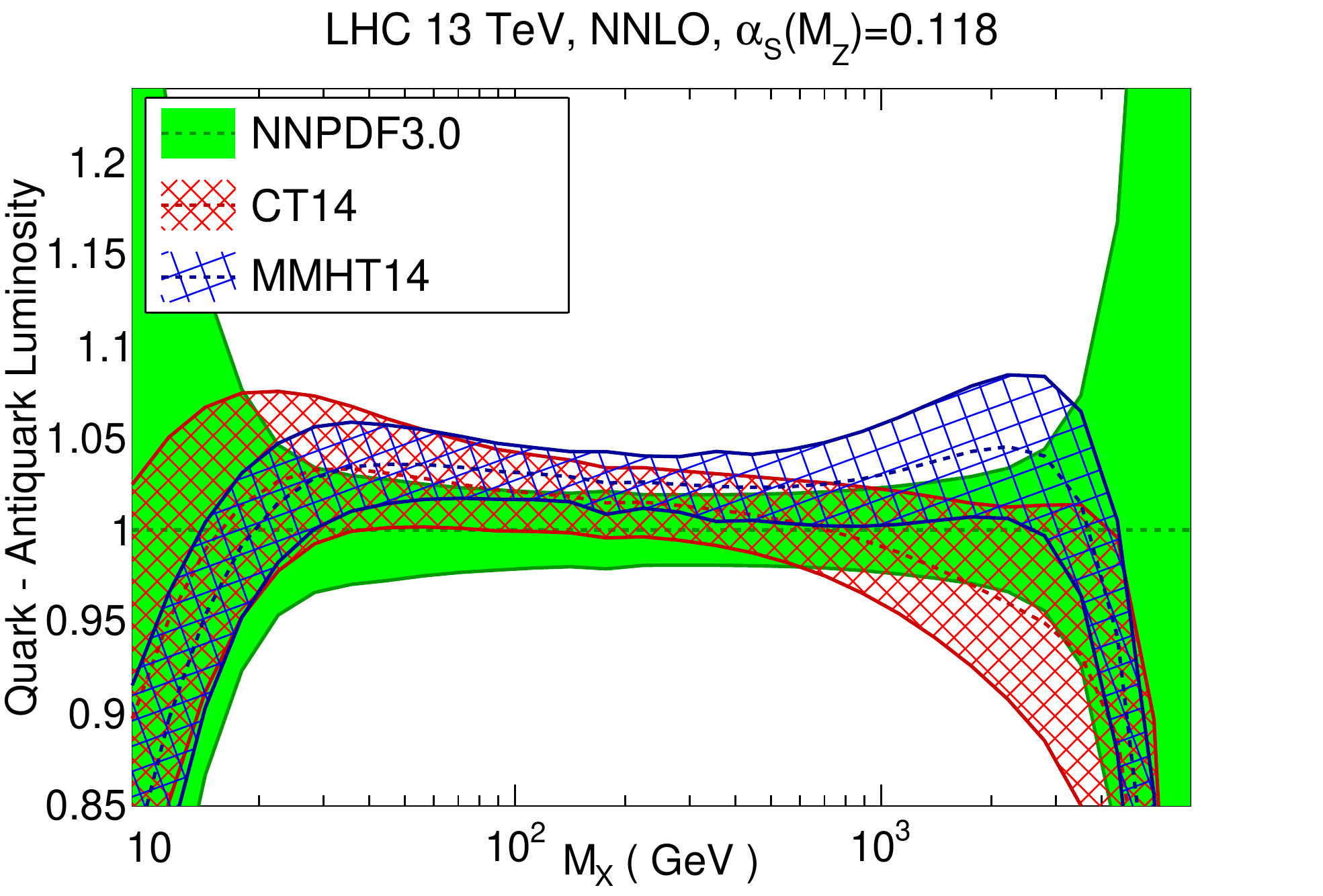}
\caption{\small Comparison of the gluon-gluon (left)
  and quark-antiquark (right) PDF luminosities
  from the NNLO NNPDF3.0, CT14 and MMHT14 global
  analyses at the LHC Run II.
}
\label{fig:lumicomp}
\end{figure}

The accurate determination of Parton Distribution Functions (PDFs)
is an essential component for a successful LHC
program~\cite{Ball:2012wy,Forte:2013wc,Rojo:2015xta}.
Recently, the three main PDF fitting collaborations have presented updates
of their global analysis: NNPDF3.0~\cite{Ball:2014uwa}, CT14~\cite{Dulat:2015mca} and
MMHT14~\cite{Harland-Lang:2014zoa}.
As compared to previous releases, an improved agreement is now found
for
some key PDF combinations such as the gluon at medium-$x$, relevant
for Higgs production in gluon fusion, though discrepancies remain in other
flavors and regions of $x$, in particular at large-$x$.
In Fig.~\ref{fig:lumicomp} we compare the NNLO gluon-gluon and quark-quark
PDF luminosities
from NNPDF3.0, CT14 and MMHT14 at a center of mass energy
of 13 TeV.

%%%%%%%%%%%%%%%%%%%%%%%%%%%%%%%%%%%%%%%%%%%%%%%%%%
%%%%%%%%%%%%%%%%%%%%%%%%%%%%%%%%%%%%%%%%%%%%%%%%%

In addition to the delivery of the updates of global PDF fits,
there has also been
intensive activity in the development of new methods for the
combination of PDF sets, motivated by ongoing updates of the PDF4LHC Working
Group prescriptions~\cite{Botje:2011sn}.
The main idea of these methods is to achieve a statistically well-defined
combination of individual PDF sets in terms of a reduced number
of Hessian eigenvectors or Monte Carlo replicas, to be used to quantify
the
overall PDF uncertainty in LHC analysis such as Higgs coupling measurements and New
Physics searches.

The starting point of these PDF
combination strategies is the Monte Carlo (MC) method~\cite{Watt:2012tq},
by means of which Hessian PDF sets are represented in terms of a MC ensemble.
Then, assuming equal likelihood for each PDF set enter in the combination,
one needs to add together the same number of replicas from each set.
The resulting combination is given in terms of a large number,
$\mathcal{O}\left( 1000\right)$,
of replicas, a number that needs to be reduced in order to streamline
its application to LHC analyses.
This can be achieved either by compressing the original MC representation
(CMC-PDFs~\cite{Carrazza:2015hva}) or by constructing a suitable Hessian representation, either by projecting
in a subspace of PDF functional forms (Meta-PDFs~\cite{Gao:2013bia}) or by using the MC replicas
themselves as basis of the linear expansion (MC2H~\cite{Carrazza:2015aoa}).
Each of these  methods is more suited for different applications,
for instance, CMC-PDFs should be more reliable for analysis where the underlying
probability distribution for the PDFs has sizable non-gaussian features,
such as New Physics searches at high invariant masses.

\begin{figure}[t]
  \centering
  \includegraphics[width=0.49\textwidth]{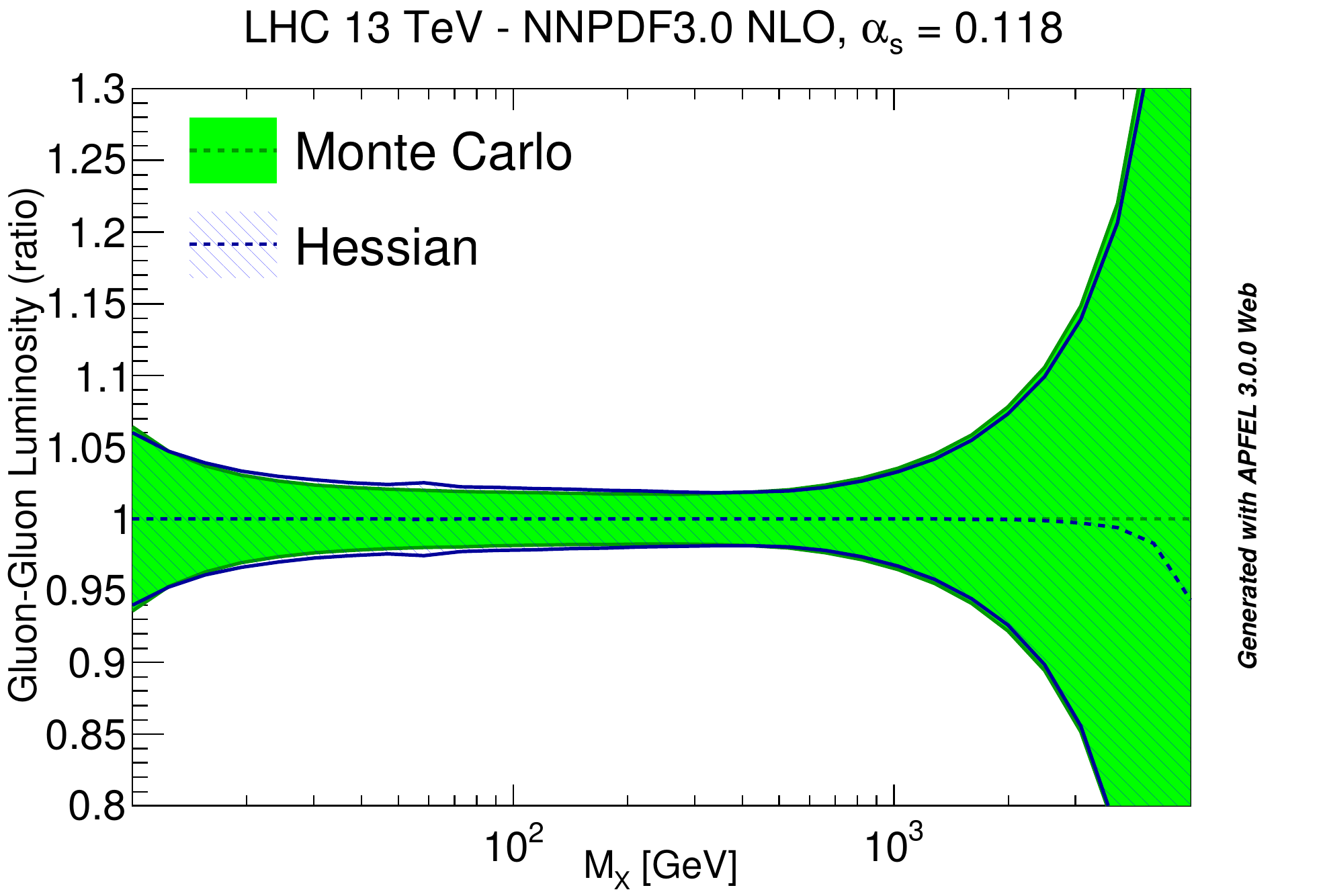}
\includegraphics[width=0.49\textwidth]{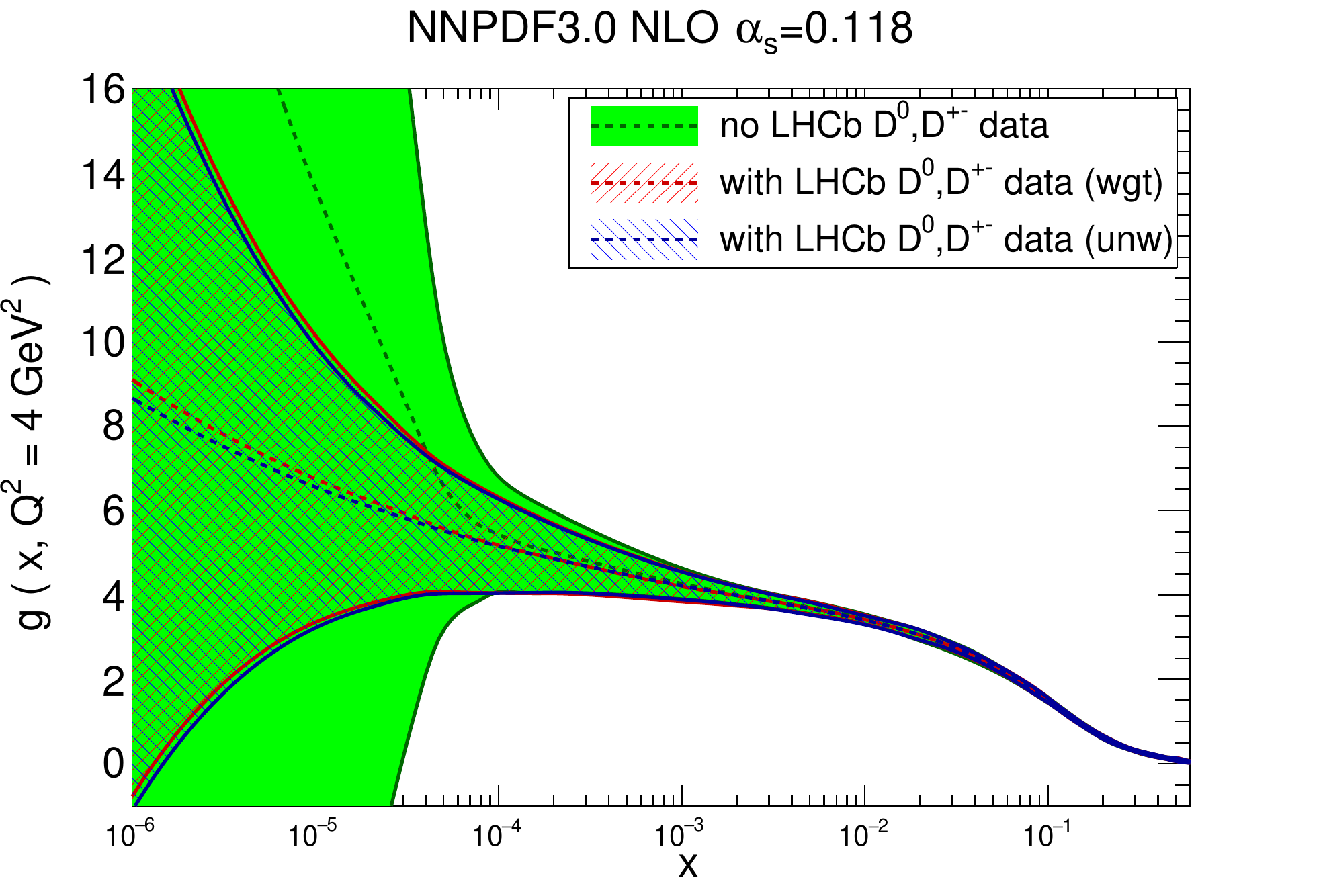}
\caption{\small Left plot:
  comparison of the native Monte Carlo and the new Hessian representations
  of NNPDF3.0 NLO.
  Right plot: constraints on the small-$x$ gluon
  PDF from the inclusion in the NNPDF3.0NLO fit of the forward charm production
  data from LHCb.
}
\label{fig:lhcbcharm}
\end{figure}

Note that as a byproduct of the development of the
MC2H method, Hessian versions of NNPDF3.0 have
also been publicly released, as illustrated in Fig.~\ref{fig:lhcbcharm}.

\paragraph{Progress in the NNPDF global analysis framework.}

Since the release of the NNPDF3.0 global fit, the main activity within the
NNPDF collaboration has been the addition of new LHC datasets that have become
available since then, as well as the addition of
the legacy HERA inclusive combination~\cite{Abramowicz:2015mha},
to be discussed in more detail below.
New measurements
that will be part of future NNPDF releases include the CMS 8 TeV double
differential distributions~\cite{CMS:2014jea},
the ATLAS 2011 inclusive jets~\cite{Aad:2014vwa}
and the LHCb
forward electroweak measurements~\cite{Aaij:2015vua}, among
several others.

In addition to these measurements, the impact on NNPDF3.0
of the LHCb data on forward charm production was recently quantified
in Ref.~\cite{Gauld:2015yia} by means of the Bayesian
reweighting method~\cite{Ball:2011gg,Ball:2010gb}.
Following a suitable normalization of the differential cross-sections,
good consistency between the LHCb data and NLO QCD theory
was found, and a substantial reduction of the gluon PDF uncertainties at
small-$x$ (beyond the HERA coverage) was obtained, as can be seen
in Fig.~\ref{fig:lhcbcharm}.
These results, consistent with a similar {\tt HERAfitter}
study~\cite{Zenaiev:2015rfa}, are a crucial input to provide robust predictions for
the charm-induced
neutrino flux at neutrino telescopes like IceCube, which is the dominant
background to astrophysical neutrinos.

%%%%%%%%%%%%%%%%%%%%%%%%%%%%%%%%%%%%%%%%%%%%%%%%%%
%%%%%%%%%%%%%%%%%%%%%%%%%%%%%%%%%%%%%%%%%%%%%%%%%

Another important recent development
in the NNPDF framework
is the  release, for the first time, of a global PDF fit which goes beyond
fixed-order QCD perturbation theory and
incorporates the effects of
soft-gluon (threshold) resummation~\cite{Bonvini:2015ira}.
The motivation for these resummed PDFs was
provided by the availability of state-of-the-art
calculations where fixed order NLO or NNLO QCD was matched to NLL or NNLL threshold
resummation, in processes  like Higgs, top quark pair production or supersymmetric pair
production: these calculations are theoretically inconsistent since up to know they had
to rely on fixed-order (rather than resummed) PDFs.

\begin{figure}[t]
\centering
\includegraphics[width=0.49\textwidth]{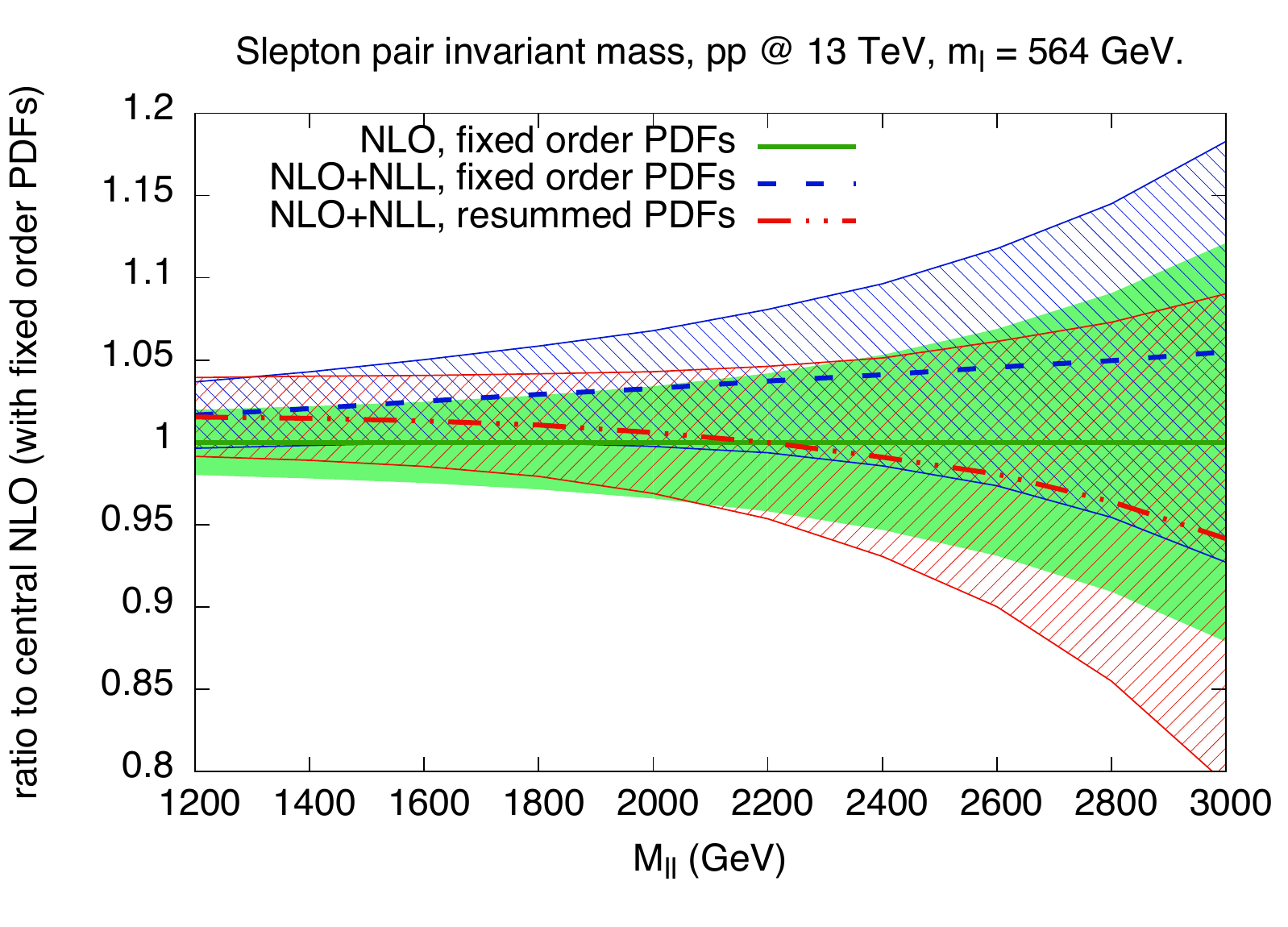}
\includegraphics[width=0.49\textwidth]{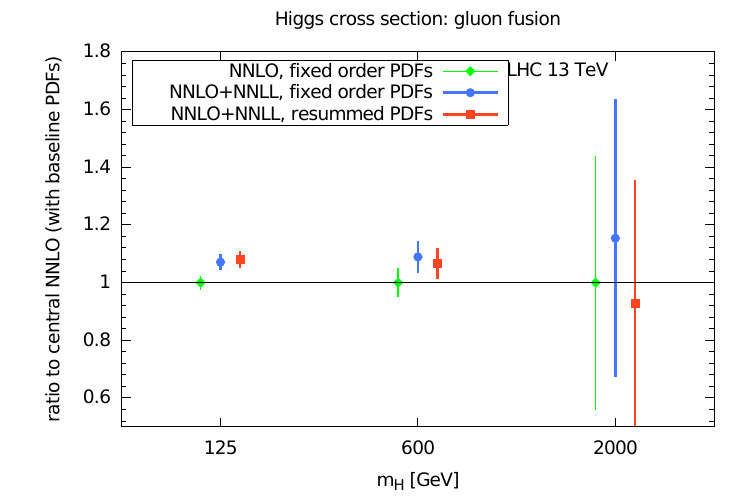}
\caption{\small Comparison of fixed-order and resummed calculations using
  either fixed-order or resummed PDFs, for slepton pair production (left)
  and for light and heavy Higgs production (right) at the LHC 13 TeV.
}
\label{fig:resummedpheno}
\end{figure}

The main outcome of this study was precisely to emphasize the crucial
importance of the consistent use of resummed PDFs with resummed partonic cross-sections.
This point is illustrated in
Fig.~\ref{fig:resummedpheno} with two examples of phenomenological relevance:
slepton pair production at high invariant masses,
and Higgs production, both SM and for a heavy Higgs.
We find that, at TeV scales,
the consistent use of resummed PDFs with resummed calculations brings the
overall result closer to its fixed-order counterpart,
while using resummed calculations with fixed-order PDFs overestimates
the overall impact of the resummation.
At moderate invariant masses, around the electroweak scale, resummed
PDFs reduce to their fixed-order counterparts, and thus the overall
effect of the resummation arises only from the partonic cross-sections.

%%%%%%%%%%%%%%%%%%%%%%%%%%%%%%%%%%%%%%%%%%%%%%%%%%
%%%%%%%%%%%%%%%%%%%%%%%%%%%%%%%%%%%%%%%%%%%%%%%%%%
%%%%%%%%%%%%%%%%%%%%%%%%%%%%%%%%%%%%%%%%%%%%%%%%%
%%%%%%%%%%%%%%%%%%%%%%%%%%%%%%%%%%%%%%%%%%%%%%%%%%

Other ongoing developments in NNPDF include fits with a fitted charm
PDF (for which a suitable modification of the FONLL general-mass
VFN scheme~\cite{Forte:2010ta}
is required), updated QED fits~\cite{Ball:2013hta} and NNPDF fits with theoretical
uncertainties, both from scale variations and from heavy quark mass
variations.
From the technical point of view, we have completed the
interfacing of the {\tt APFEL} program~\cite{Bertone:2013vaa}
with the
NNPDF fitting code, which allows to streamline the study
of the impact, at the PDF level, of new theory settings.

\paragraph{PDF impact of the legacy HERA inclusive combination.}

\begin{figure}[t]
\centering
\includegraphics[width=0.49\textwidth]{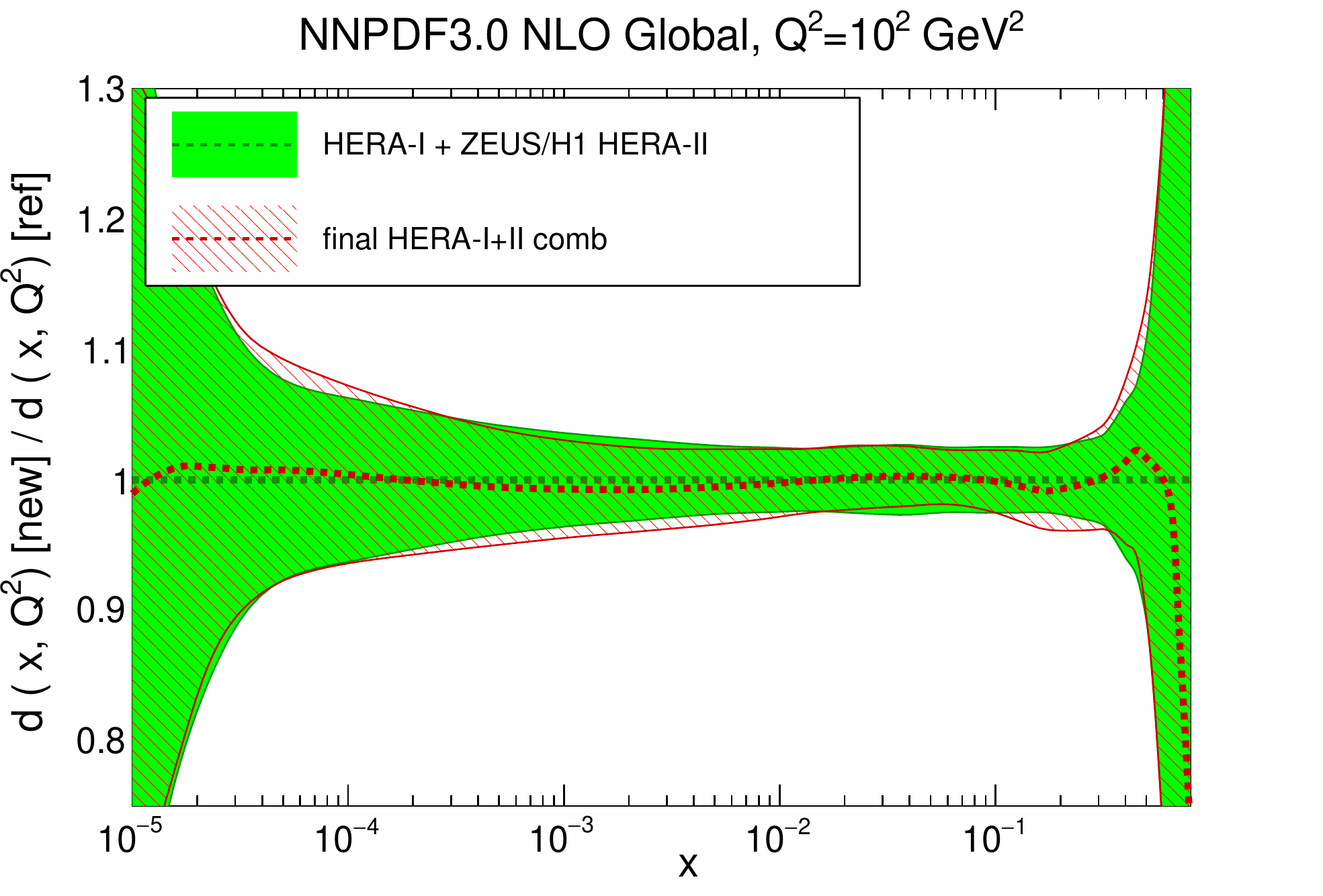}
\includegraphics[width=0.49\textwidth]{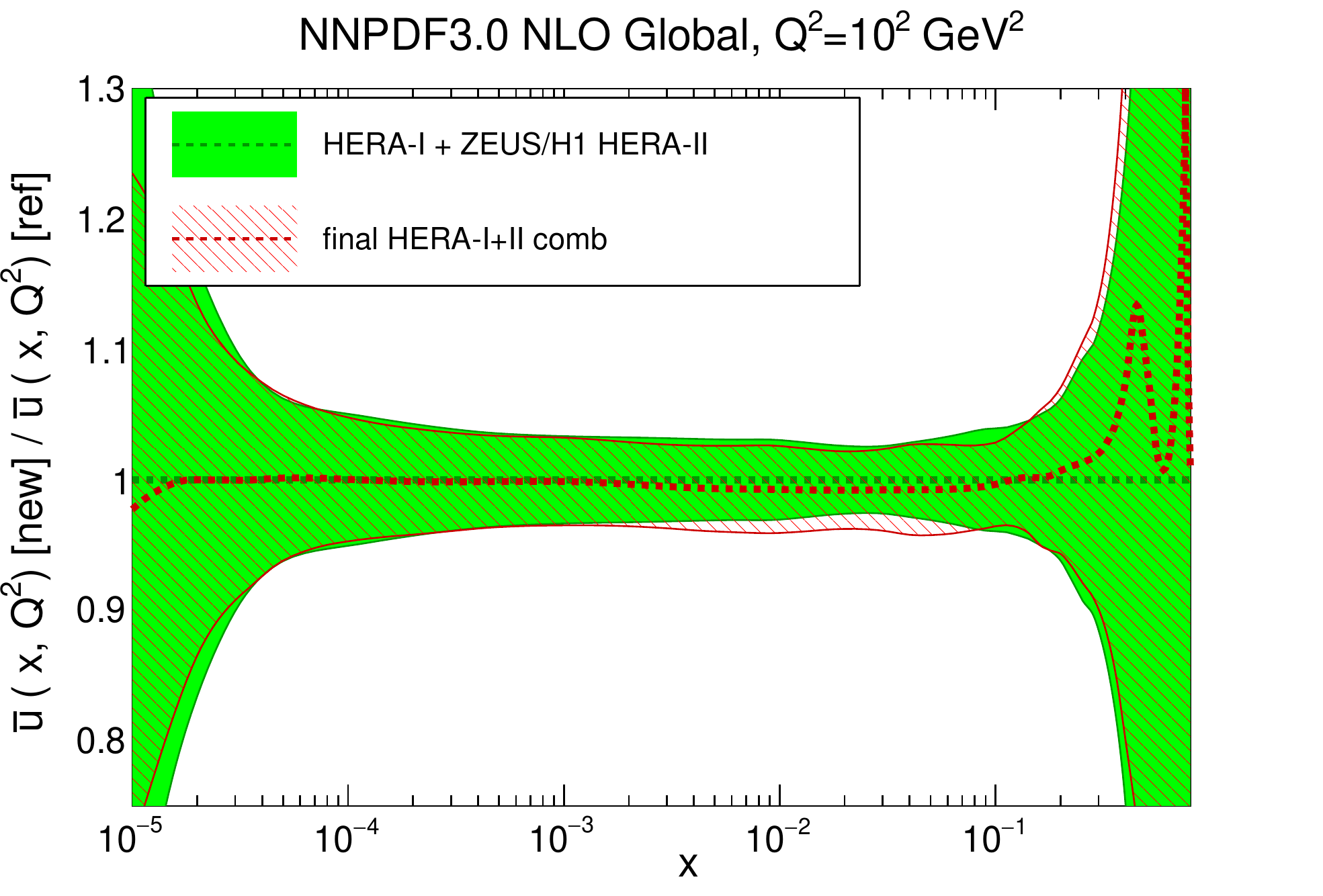}
\caption{\small Comparison of the down quark $d(x,Q^2)$ (left plot) and
  the anti-up quark $\bar{u}$ PDFs (right plot) in the baseline NNPDF3.0 NLO fit
  and in the variant where all inclusive HERA data has been replaced
  by the legacy combination.
  The comparison is performed at $Q^2=100$ GeV$^2$.
}
\label{fig:heraGlobal}
\end{figure}

The H1 and ZEUS collaborations have recently presented the legacy combination
of all HERA inclusive structure function data from Runs I and
II~\cite{Abramowicz:2015mha}.
This combined dataset supersedes both the HERA-I combination~\cite{Aaron:2009aa} and the separate
HERA-II measurements from H1 and ZEUS.
In the NNPDF3.0 global fit, both the HERA-I data and the inclusive measurements
from HERA-II were already included, so one expected that the replacement
of these with the legacy combination would have a very moderate effect.
To verify if this expectation is borne out, we have produced NLO and NNLO
versions of NNPDF3.0 with the HERA legacy combination replacing
the corresponding inclusive measurements.
The comparison is shown in Fig.~\ref{fig:heraGlobal}.
Indeed we verify that the impact of the final HERA combination on the
NNPDF3.0 fits
is minimal, both for central values and for PDF uncertainties.
%

%%%%%%%%%%%%%%%%%%%%%
%%%%%%%%%%%%%%%%%%%%

\begin{figure}[t]
\centering
\includegraphics[width=0.49\textwidth]{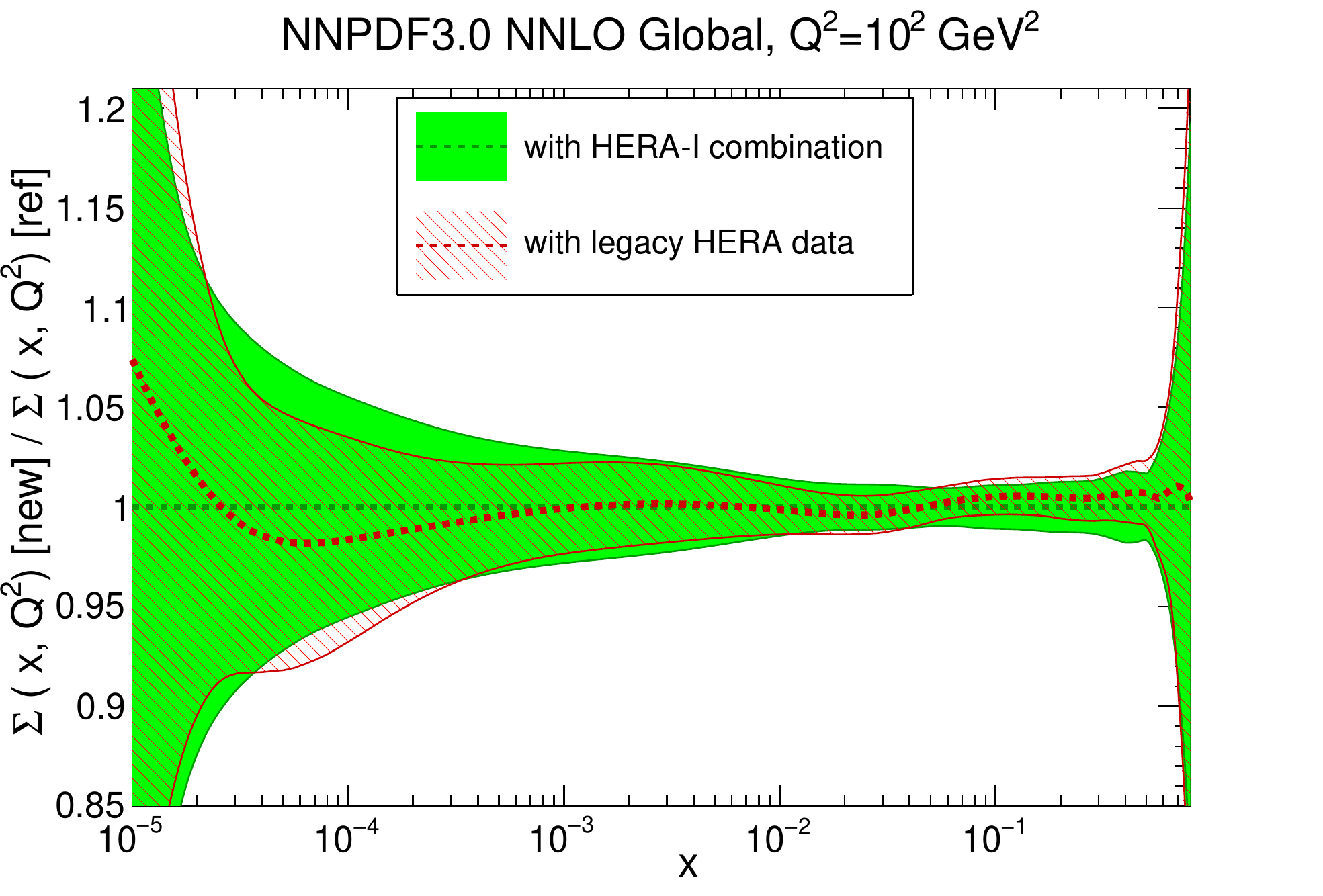}
\includegraphics[width=0.49\textwidth]{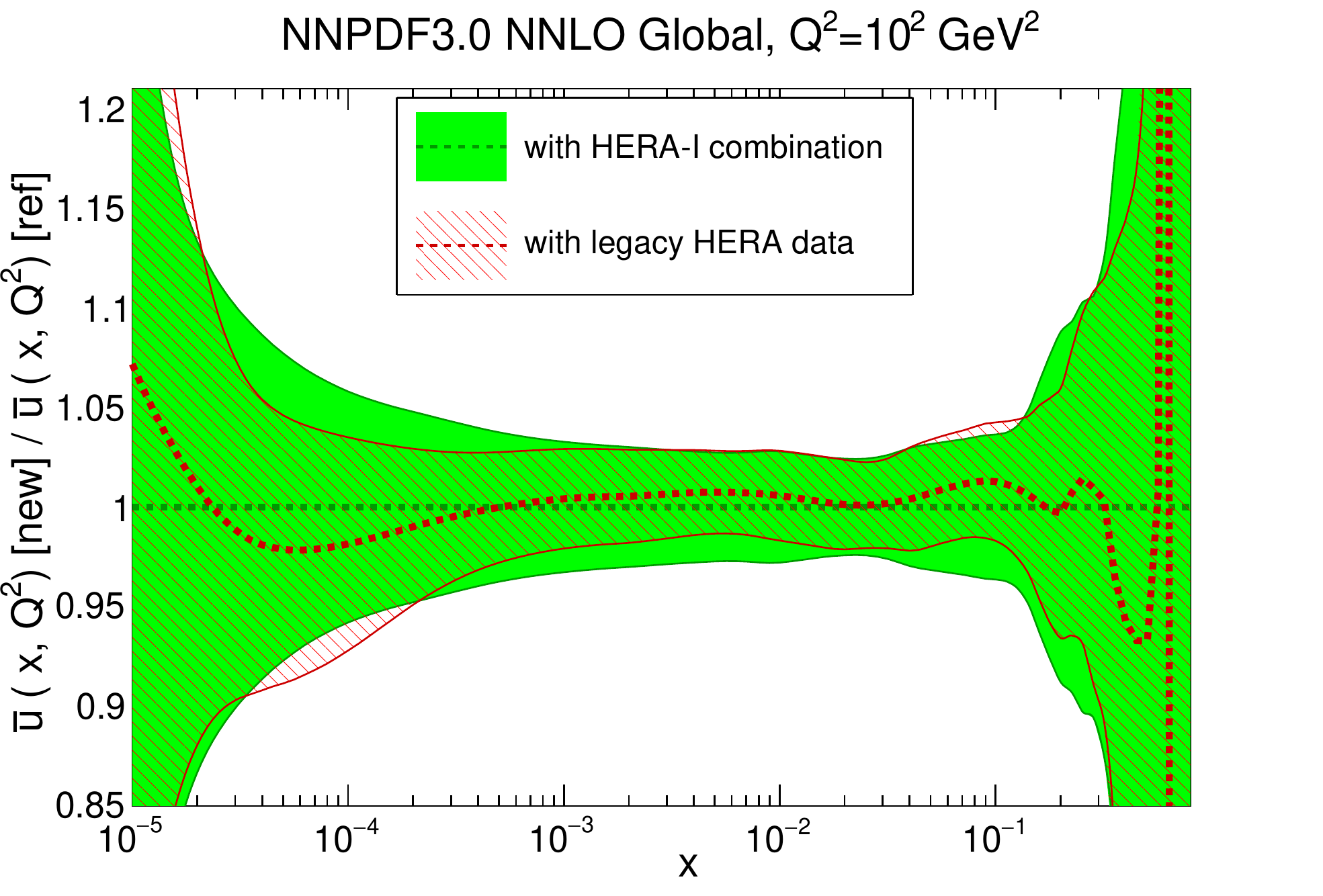}
\caption{\small \label{fig:heraGlobal2}
  Same as Fig.~4 now comparing the total quark singlet  $\Sigma(x,Q^2)$
  and the anti-up quark $\bar{u}(x,Q^2)$ PDFs from the NNPDF3.0 NNLO
  global fit with the legacy HERA dataset,
  with a global fit
  without any HERA-II inclusive data.}
\end{figure}

The above conclusion is perhaps not completely surprising, since
we already verified that, in the NNPDF3.0 global analysis, the impact
of the individual HERA-II inclusive measurements from H1 and ZEUS
was moderate to begin with~\cite{Ball:2014uwa}, though providing
some useful constraints on quarks at medium and large $x$.
The situation is unchanged with the availability of the legacy
HERA combination, as illustrated in
Fig.~\ref{fig:heraGlobal2}, where we compare the global fit,
including the final HERA dataset, with the same fit, but now
including only the HERA-I data.
We observe that the central values are essentially unchanged in the two fits,
with a moderate reduction of the PDF uncertainties on the quark PDFs at intermediate
values of $x$.
Therefore, we conclude that the addition of the HERA-II inclusive
data into a global PDF fit
that already includes the HERA-I combination, while leaving unchanged
the central values, helps in reducing the PDF uncertainties
of quarks, and that the impact of the combination {\it per se} is negligible.
Let us 
mention that more detailed studies of the impact of the legacy HERA combination
in the NNPDF global fit will be presented in a future publication.
The impact of this dataset in the MMHT14 framework has recently
been explored in Ref.~\cite{Thorne:2015caa}.

%%%%%%%%%%%%%%%%%%%%%
%%%%%%%%%%%%%%%%%%%%

\begin{figure}[t]
\centering
\includegraphics[width=0.49\textwidth]{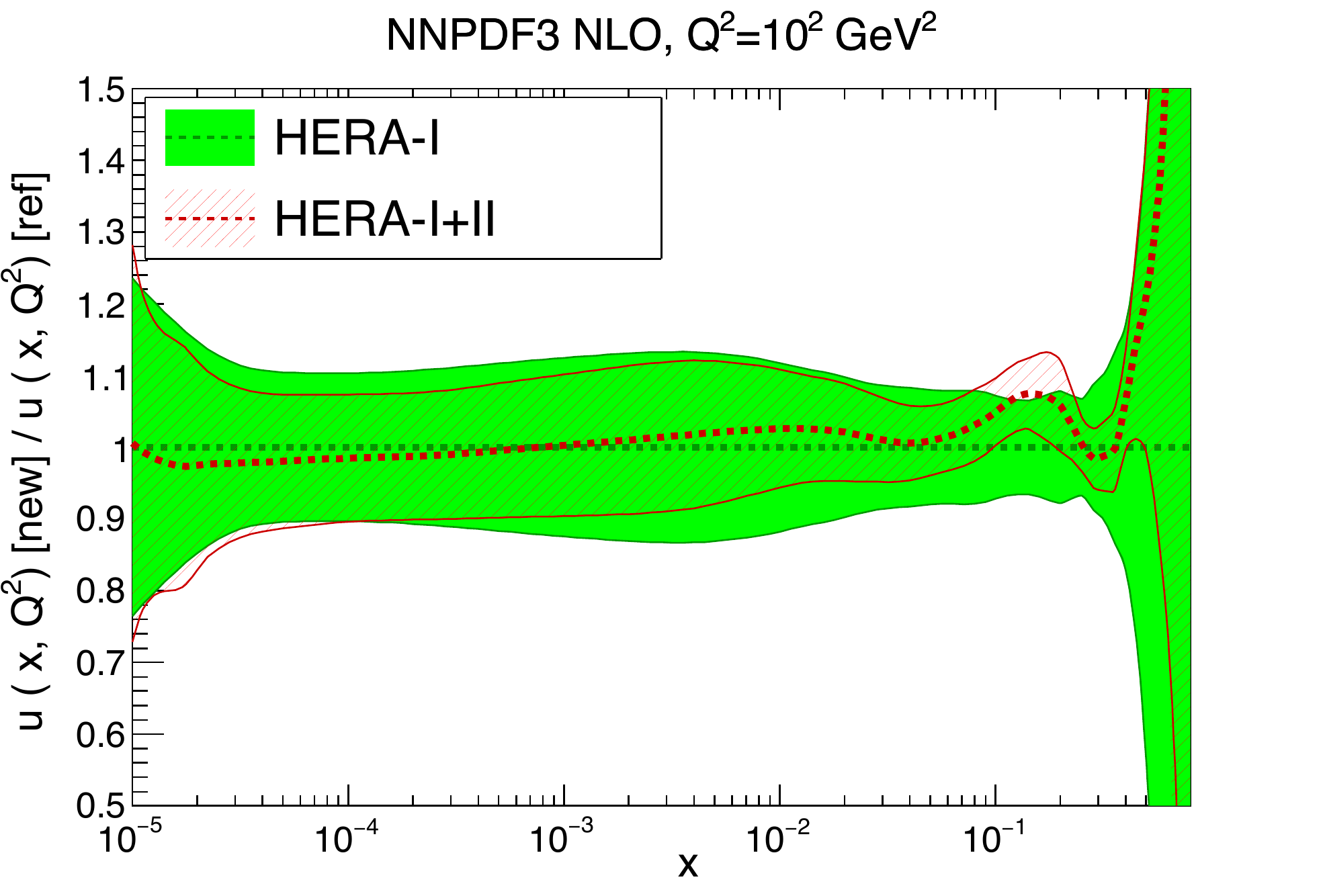}
\includegraphics[width=0.49\textwidth]{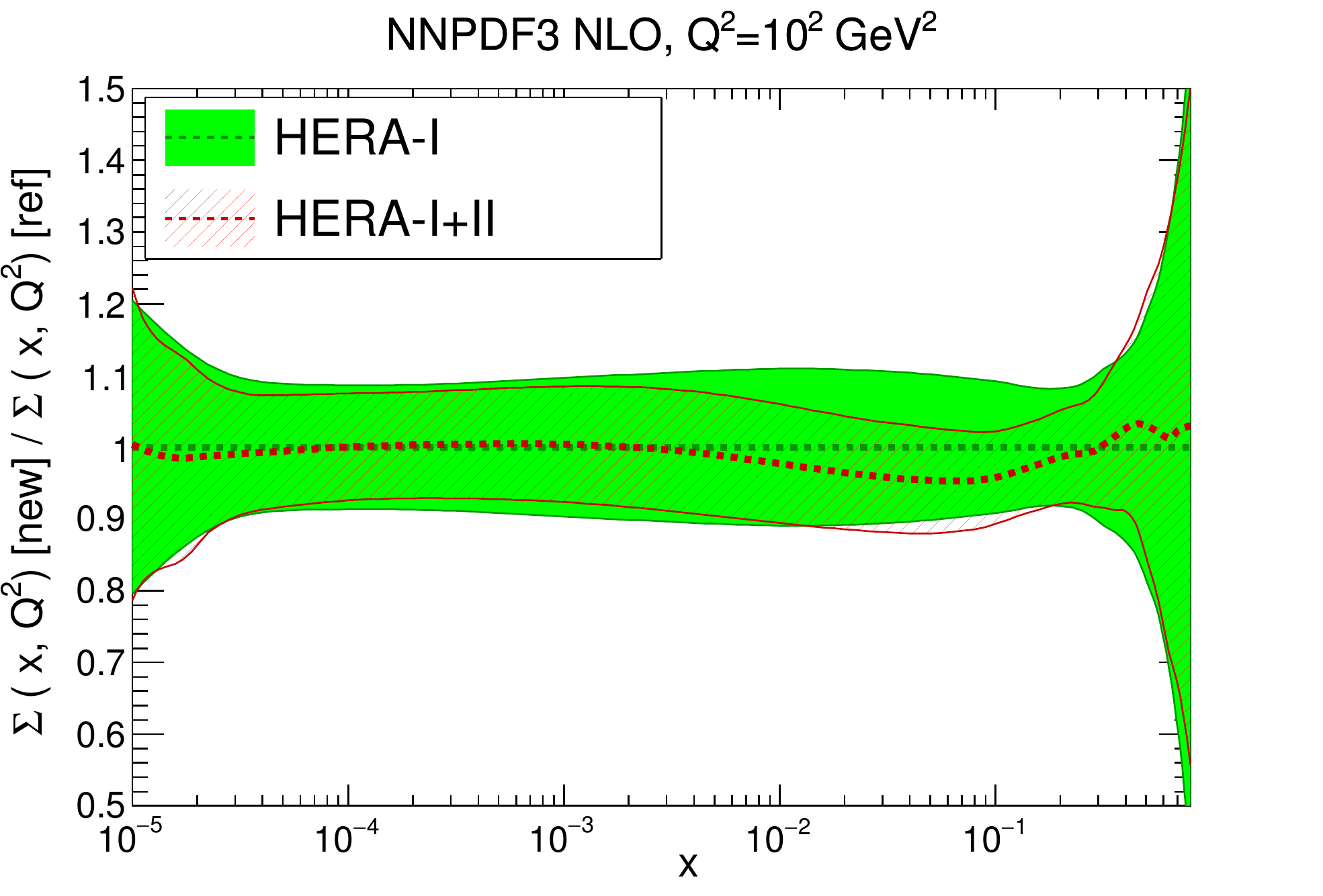}
\caption{\small Comparison of the up quark (left plot) and total
  quark singlet (right plot) in a HERA-I only fit and in a PDF fit based
on the final HERA-I+II combination.}
\label{fig:heraonly}
\end{figure}

In the same publication that presented the HERA legacy
combination~\cite{Abramowicz:2015mha}, the HERAPDF2.0 
analysis was also introduced, and
a substantial reduction of PDF uncertainties as compared to the previous
HERAPDF1.5 was reported.
To explore the impact that the legacy HERA combination has as compared
to a HERA-I only fit in the NNPDF framework, we have produced variants of NNPDF3.0 based only
on either HERA-I data or on the legacy HERA combination.
In Fig.~\ref{fig:heraonly} we show the comparison of the up quark (left plot) and total
quark singlet (right plot) in a HERA-I only fit and in a PDF fit including
only the final HERA-I+II combination.
  We now observe that the impact of the new HERA data is larger than in the global
  fit, in particular in what concerns quark flavor separation at medium and
  large-$x$.
  This is consistent with the results of the HERAPDF2.0 paper.

%%%%%%%%%%%%%%%%%%%%%
%%%%%%%%%%%%%%%%%%%%

\begin{figure}[t]
\centering
\includegraphics[width=0.70\textwidth]{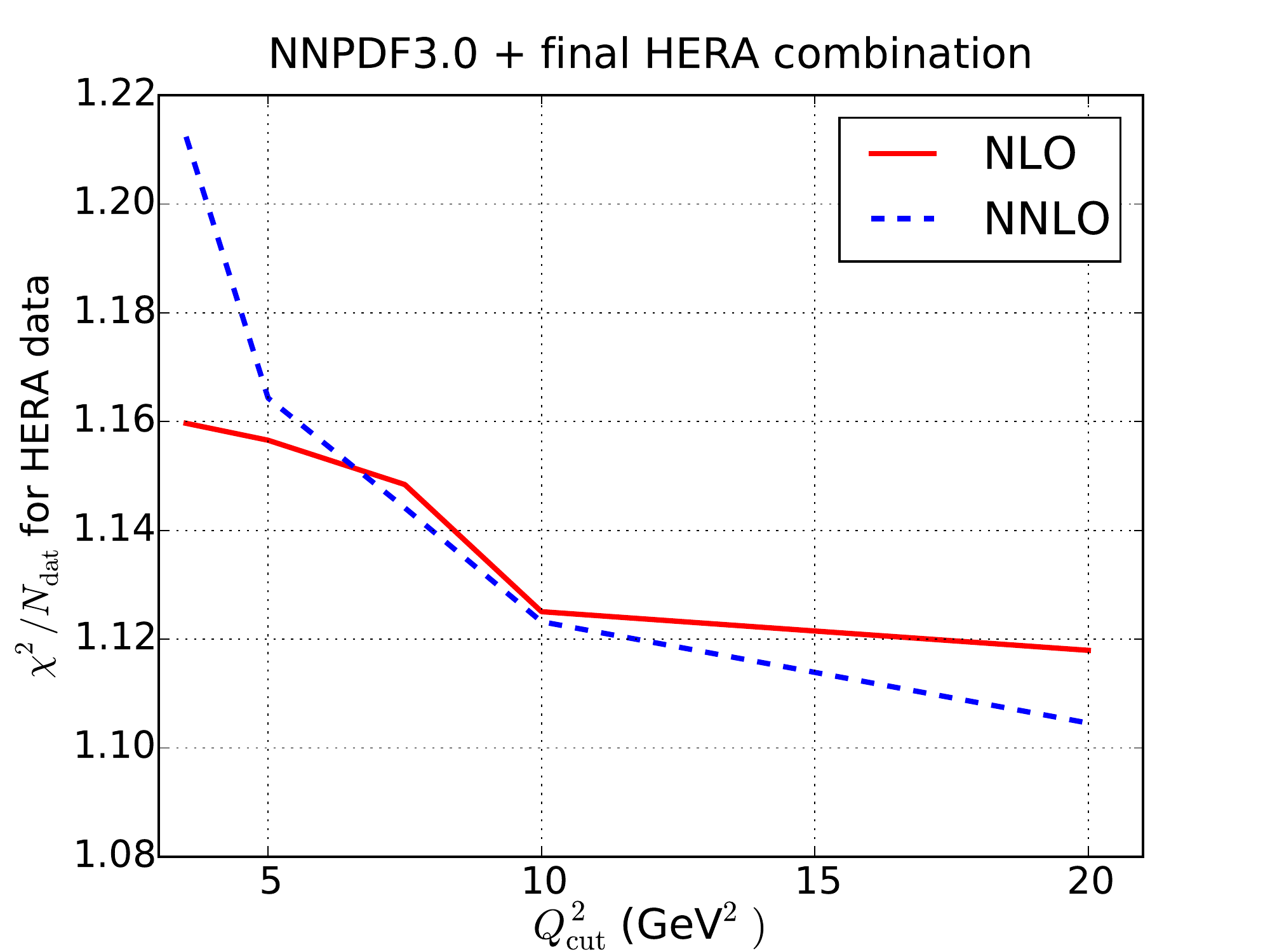}
\caption{\small The value of $\chi^2/N_{\rm dat}$ for the legacy HERA combination
  in the variants of the NNPDF3.0 fits with different values of $Q^2_{\rm cut}$,
for the NLO and NNLO fits.}
\label{fig:heraplot}
\end{figure}

The HERAPDF2.0 analysis also reported~\cite{Abramowicz:2015mha}
a sizable dependence of the fit results and the data $\chi^2$ values
with respect to the choice for
the minimum value of $Q^2$, denoted by $Q^2_{\rm min}$, included in the analysis.
Such instability, if confirmed by other groups,
could have different origins, like an inadequacy of the theory
used for the fit, for example if small-$x$ (BFKL) resummation~\cite{Altarelli:2008aj}
is needed to describe
the precise inclusive HERA data at low-$x$ and low-$Q^2$.
To verify this
observation, we have produced variants of the NNPDF3.0 global fit,
including the HERA legacy combination,
for different values of $Q^2_{\rm min}$.
The results of this study are summarized in
Fig.~\ref{fig:heraplot}, where we show for the NLO and NNLO fits
the value of $\chi^2/N_{\rm dat}$  as a function of $Q^2_{\rm cut}$.

%%%%%%%%%%%%%%%%%%%%%
%%%%%%%%%%%%%%%%%%%%

From Fig.~\ref{fig:heraplot} we see that
also in NNPDF3.0 we observe that the $\chi^2/N_{\rm dat}$ of the HERA data decreases
quite rapidly as $Q^2_{\rm cut}$ is increased, more at NNLO than at NLO.
This effect disappears for $Q^2_{\rm min}\ge 10$ GeV$^2$, for which the $\chi^2$ profiles
essentially flatten out.
Interestingly, for $Q^2_{\rm min}\ge 5$ GeV$^2$ we see that the quality of the NNLO fit
is essentially the same or better than for the NLO fit.
These results are consistent with the possibility of large unresummed small-$x$ logarithms,
though there are also alternative explanations, such as an
possible internal tension between the HERA data at small-$Q^2$ and the rest
of the dataset.

\begin{figure}[t]
\centering
\includegraphics[width=0.49\textwidth]{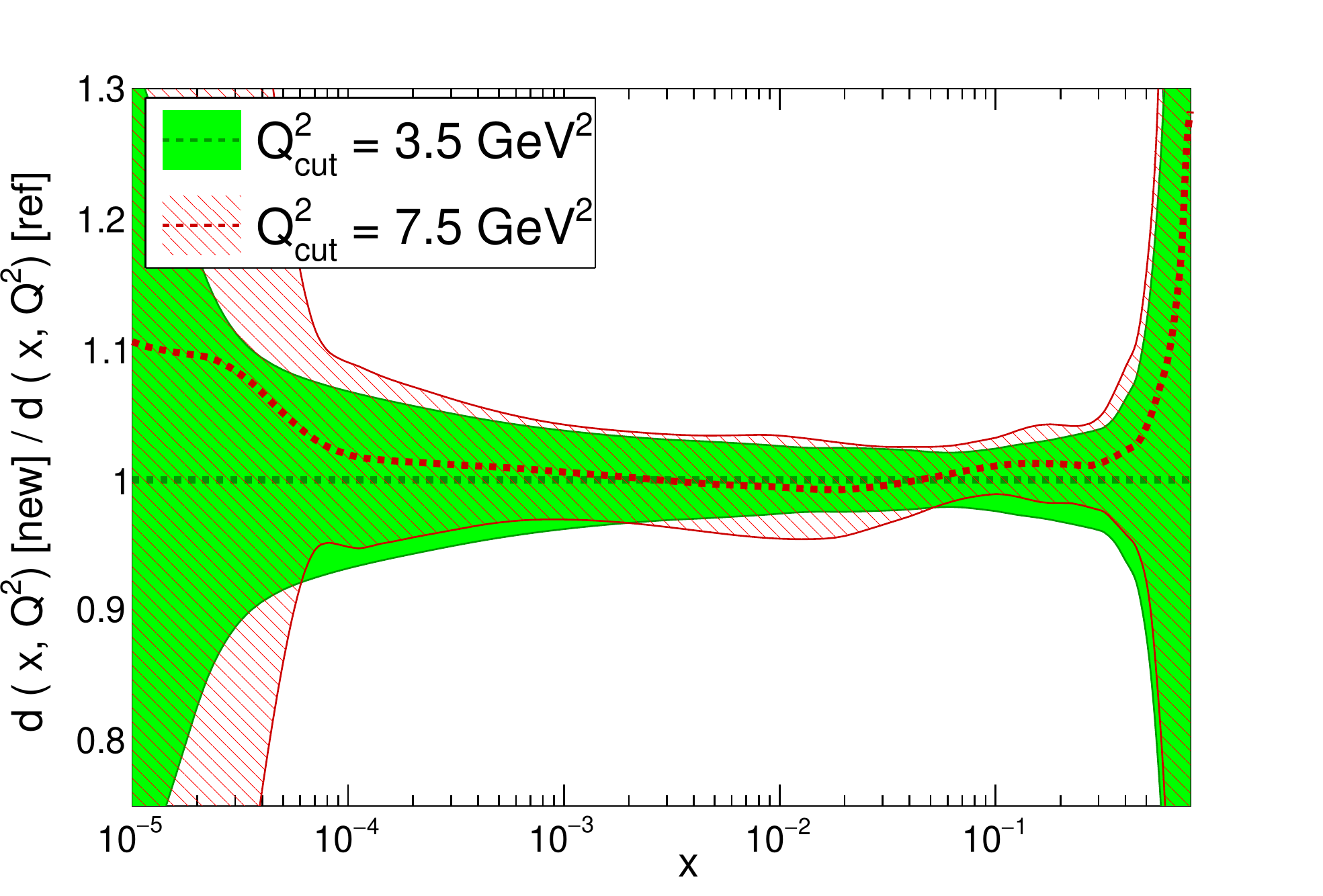}
\includegraphics[width=0.49\textwidth]{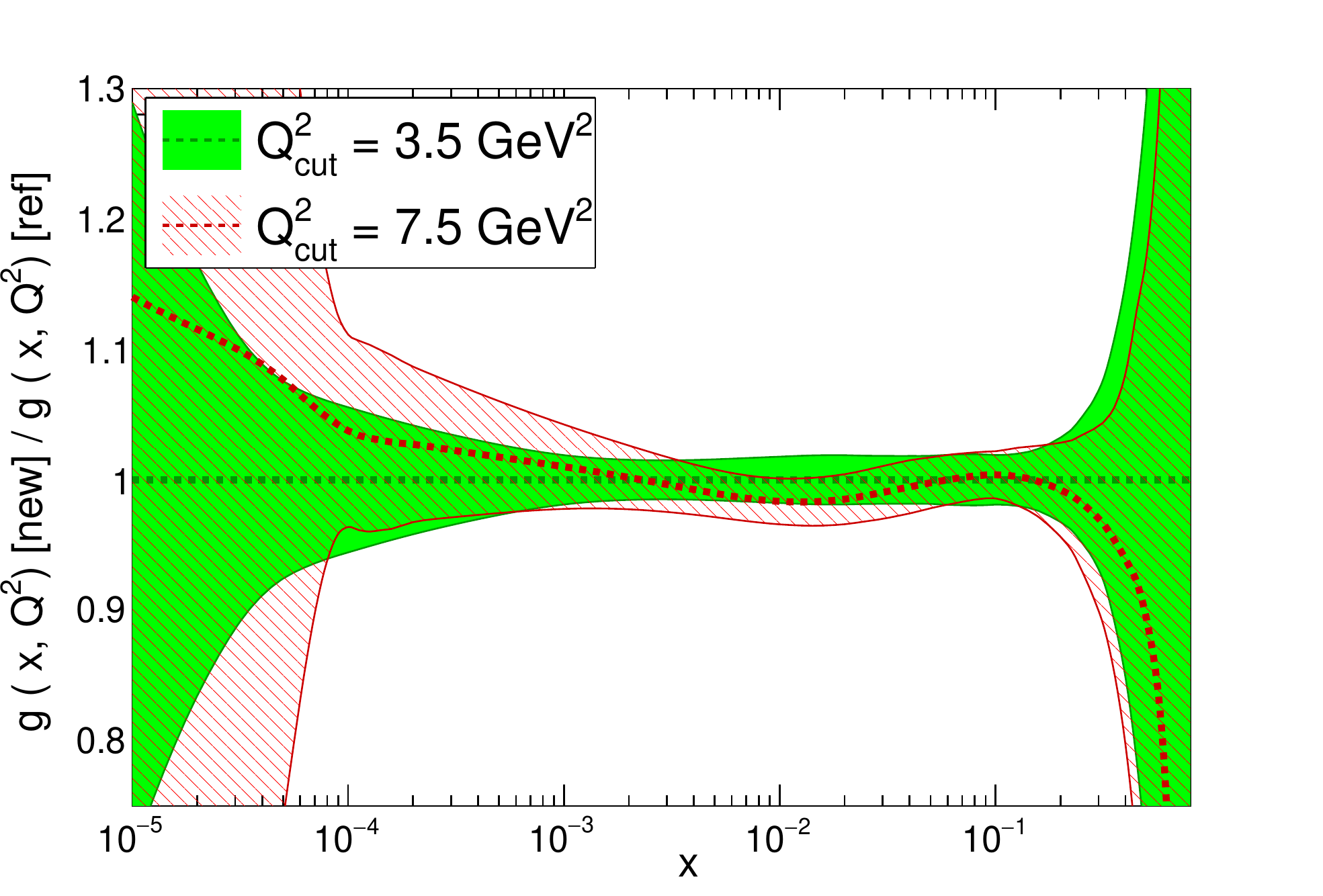}
\caption{\small Comparison of variants of the NNPDF3.0 NLO global
  fit with two different values of $Q^2_{\rm cut}$, namely 3.5 GeV$^2$
  (the default) and 7.5 GeV$^2$, for the down quark
  (left plot) and the gluon PDF (right plot).
}
\label{fig:heraQ2cut}
\end{figure}

Whatever the origin of this dependence with $Q^2_{\rm min}$ is, it is reassuring that its impact
on mainstream LHC physics, which typically probe PDFs at medium and large-$x$ and at large
$Q^2$, is moderate.
In Fig.~\ref{fig:heraQ2cut} we compare the NNPDF3.0 fits (with the legacy combination) for two values
of $Q^2_{\rm min}$, namely 3.5 GeV$^2$
(which is the default in NNPDF3.0) and 7.5 GeV$^2$.
As we can see, while of course the PDF uncertainties at low-$x$ increase strongly,
for $x\gsim 10^{-3}$ the impact of removing low $Q^2$ data is small.
The same holds true for other PDF combinations.

%%%%%%%%%%%%%%%%%%%%%
%%%%%%%%%%%%%%%%%%%%

However, it is also important to note that in order to probe the possible
effect of BFKL-like unresummed logs, it is more efficient~\cite{Caola:2009iy}
to use
a kinematical cut of the form
\begin{equation}
Q^2 \ge A_{\rm cut} x^{-\lambda} \, ,
\end{equation}
with $\lambda\sim 0.3$ and $A_{\rm cut}$  a parameter that is varied,
analogously to $Q^2_{\rm cut}$.
This cut removes only small-$x$, small-$Q^2$ data but not small-$Q^2$
and large-$x$ data that is certainly unaffected by small-$x$ resummation
(and where other effects, like TMCs and higher twists, might play a role).
In any case, only a complete global analysis
where NLO and NNLO fixed-order calculations are supplemented
with small-$x$ resummation should be able to definitely elucidate
the origin of these effect.
Such fit in the
NNPDF framework is underway.

\paragraph{Acknowledgments}
We are grateful to  R.~Thorne, A.~Cooper-Sarkar and V.~Radescu
for many useful discussions on the impact on
PDFs of the HERA legacy combination.
This work has been supported by an STFC Rutherford Fellowship
and Grant ST/K005227/1 and ST/M003787/1, and
by an European Research Council Starting Grant "PDF4BSM".

\end{document}